\documentclass[%
 reprint,
superscriptaddress,
 amsmath,amssymb,
 aps,
]{revtex4-2}

\usepackage{graphicx}% Include figure files
\usepackage{dcolumn}% Align table columns on decimal point
\usepackage{bm}% bold math
\usepackage{amsmath}
\usepackage{color} %textcolor
\usepackage{soul}%strikeout
\usepackage[normalem]{ulem}
\usepackage{hyperref}% add hypertext capabilities
\usepackage[english]{babel}
\usepackage{physics}
\usepackage{float}

\usepackage{color}
\usepackage[normalem]{ulem}

\usepackage{tikz}
\usepackage{tikzscale} 
\usetikzlibrary{positioning}
\usetikzlibrary{decorations.pathmorphing}
\usetikzlibrary{arrows, arrows.meta}
\usepackage{tabularx}
\usepackage{pgfplots}
\usepackage{amsmath}
\pgfplotsset{compat=1.18}
\begin{document}

\raggedbottom

\title{Proposal for the generation of continuous-wave vacuum ultraviolet laser light for Th-229 isomer precision spectroscopy}

\date{\today}% It is always \today, today,

\author{Qi Xiao}
\affiliation{State Key Laboratory of Low Dimensional Quantum Physics, Department of Physics, Tsinghua University, Beijing 100084, China}

\author{Gleb Penyazkov}
\affiliation{State Key Laboratory of Low Dimensional Quantum Physics, Department of Physics, Tsinghua University, Beijing 100084, China}

\author{Ruihan Yu}
\affiliation{State Key Laboratory of Low Dimensional Quantum Physics, Department of Physics, Tsinghua University, Beijing 100084, China}

\author{Beichen Huang}
\affiliation{State Key Laboratory of Low Dimensional Quantum Physics, Department of Physics, Tsinghua University, Beijing 100084, China}

\author{Jiatong Li}
\affiliation{State Key Laboratory of Low Dimensional Quantum Physics, Department of Physics, Tsinghua University, Beijing 100084, China}

\author{Juanlang Shi}
\affiliation{State Key Laboratory of Low Dimensional Quantum Physics, Department of Physics, Tsinghua University, Beijing 100084, China}

\author{Yanmei Yu}
\email{ymyu@aphy.iphy.ac.cn}
\affiliation{Beijing National Laboratory for Condensed Matter Physics, Institute of Physics, Chinese Academy of Sciences, Beijing, 100190, China}
\affiliation{University of Chinese Academy of Sciences, 100049 Beijing, China}

\author{Yuxiang Mo}
\email{ymo@mail.tsinghua.edu.cn}
\affiliation{State Key Laboratory of Low Dimensional Quantum Physics, Department of Physics, Tsinghua University, Beijing 100084, China}

\author{Shiqian Ding}
\email{dingshq@mail.tsinghua.edu.cn}
\affiliation{State Key Laboratory of Low Dimensional Quantum Physics, Department of Physics, Tsinghua University, Beijing 100084, China}

\begin{abstract}
We propose to generate continuous-wave vacuum ultraviolet (VUV) laser light at 148.4~nm using four-wave mixing in cadmium vapor for precision spectroscopy of the Th-229 isomer transition. Due to the large transition matrix elements of cadmium, the readily accessible wavelengths for the incident laser beams, and the high coherence of the four-wave mixing process, over 30 $\mu$W of VUV power can be generated with a narrow linewidth. This development paves the way for coherently driving the Th-229 isomer transition and developing the nuclear optical clock. 
\end{abstract}

\maketitle

The invention of coherent laser sources has revolutionized the field of atomic and molecular physics.
The success of these applications can be attributed to the ability to drive transitions in the electron shell using lasers.
Expanding these well-established technologies to manipulate nuclear quantum states holds great promise. 
However, coherent control of nuclear transitions has been challenging~\cite{CoherentNuclear2006,CoherentNuclear2021,Sc45} due to the significant energy gap between the nuclear transitions and the photon energy of coherent laser radiation.
A nuclear transition involves large changes of both Coulomb interaction energy and strong interaction energy that are normally a few 10$^5$-10$^6$ electron volts. 
A fortuitous cancellation of the changes of these two terms occurs in the well-known Th-229 isomer transition~\cite{2007HayesCoulomb,2009FlambaumCoulomb}, resulting in a nuclear excited state with the lowest known energy that lies only 8.4 eV above the ground state~\cite{1976Energy,1990Energy,1994Energy,2007Energy,2019EnergyMunich,2019EnergyJapan,2019EnergyJapanPRL,2020Energy,2020Energy,2022EnergyCERN,LaserSpecSchumm,LaserSpecHudson}. The corresponding transition falls within the range of typical electronic transitions, and is the only nuclear transition that can be driven using currently available laser sources. 
This offers a unique opportunity to bridge the fields of atomic and nuclear physics and potentially apply quantum technologies developed over the past few decades to nuclei. 

One remarkable frontier is the development of a nuclear optical clock~\cite{ClockPeik,ClockKuzmich}. Due to their much smaller size and multipole moments compared to atoms, nuclei exhibit greater robustness in the frequency of nuclear transitions against perturbing external fields.
This characteristic can be employed to alleviate the significant effort required to control ambient fields near an atomic optical clock. The use of the Th-229 isomer transition in a nuclear clock is further motivated by its remarkable sensitivity to temporal variations in fundamental constants, such as the fine-structure constant~\cite{FlambaumVariation,2007HayesCoulomb,2009Flambaum,Th2IsomerSpec,Th3JapanNature}, and  its application in the search for ultralight dark matter~\cite{DarkMatter2015,DarkMatter2023}. 

Another salient feature of nucleus-based experiments is the doping of a large number of atoms into a solid material while generically keeping the nuclear structure intact~\cite{ClockPeik,CrystalHudson,CrystalSchumm}. 
This is in contrast to current atomic-shell-based quantum systems, where valence electrons of suspended atoms in the vacuum play a central role, and doping would eliminate the relevant underlying quantum states.
Hosting nuclei with a solid material in the Lamb-Dick regime may substantially reduce the required experimental efforts, particularly the challenging tasks of laser cooling and trapping.
Additionally, the ability to control a macroscopic quantity of nuclear systems can largely reduce the interrogation time for precision measurements.

Driven by these compelling prospects~\cite{ThoriumReviewThirolf,ThoriumReviewNuClock,ThoriumReviewPeik}, several methods have been employed to measure the energy of the Th-229 isomer transition~\cite{1976Energy,1990Energy,1994Energy,2007Energy,2019EnergyMunich,2019EnergyJapan,2019EnergyJapanPRL,2020Energy,2022EnergyCERN,LaserSpecSchumm,LaserSpecHudson}, including indirect $\gamma$ spectroscopy~\cite{2007Energy,2019EnergyJapanPRL,2020Energy}, internal-conversion electron spectroscopy~\cite{LarsDirect,2019EnergyMunich}, and vacuum ultraviolet (VUV) spectroscopy of radiative fluorescence light~\cite{2022EnergyCERN}. 
Very recently, laser spectroscopy has been finally achieved by two groups based on thorium-doped VUV-transparent crystals~\cite{LaserSpecSchumm,LaserSpecHudson}, and the transition wavelength is determined to be 148.4 nm. 
In both experiments, tunable pulsed VUV laser sources are produced by four-wave mixing (FWM) in xenon gas, with durations of several nanoseconds, average powers ranging from 30 to 450 $\mu$W, and linewidths of a few GHz.
The current spectroscopy uncertainty stands at approximately 3 GHz~\cite{LaserSpecSchumm,LaserSpecHudson}, primarily limited by the broad spectral linewidth associated with the pulsed nature of the VUV laser.
The radiative lifetime of the isomer state is measured to be between 2000 and 3000 s~\cite{2022EnergyCERN,LaserSpecSchumm,LaserSpecHudson,Th3JapanNature,JapanXRay}, corresponding to a linewidth of around 60 $\mu$Hz. 
This significant linewidth disparity of many orders of magnitude underscores the need for a narrow-line laser for efficient transition driving and precision spectroscopy.
A highly coherent frequency comb in the VUV region has been demonstrated via high-order harmonic generation (HHG) from an infrared comb~\cite{VUVYe2005,VUVHansch2005,VUVReview} and is being utilized in Th-229 isomer spectroscopy~\cite{LarsVUV,VUVSchumm,ThoriumReviewNuClock,ChuankunVUV}. Although the linewidth of the comb might be as narrow as kHz, a single comb line with a power of only 1 nW contributes to driving the isomer transition~\cite{ThoriumReviewNuClock,ChuankunVUV}.  

In this work, we propose to generate continuous-wave (CW) laser light at 148.4 nm using four-wave mixing scheme for precision spectroscopy of the Th-229 isomer transition.
Cadmium vapor is identified as a promising nonlinear medium capable of producing over 30 $\mu$W of VUV power.
The generated VUV laser can achieve a linewidth several orders of magnitude 
narrower than pulsed FWM sources currently used in isomer spectroscopy~\cite{PeikFWM,HudsonFWM}, 
and a power more than four orders of magnitude higher than a single comb line in state-of-the-art VUV frequency comb~\cite{ChuankunVUV}.
This approach addresses the challenge posed by the lack of an intense and narrow-linewidth laser near the Th-229 isomer transition~\cite{ThirolfViewpoint}, enabling coherent driving of nuclear Rabi oscillations.

The FWM process relies on the third-order nonlinear susceptibility, which is the lowest-order nonlinearity observed in an isotropic medium.
It has long been employed to generate VUV light~\cite{TwoResonance,VidalReview} that otherwise is challenging to produce for spectroscopy in this wavelength region. The nonlinear process necessitates high incident laser intensities, generally resulting in the generated VUV light being pulsed. 
A notable exception is the production of CW Lyman-$\alpha$ radiation at 121.6~nm using FWM in mercury vapor~\cite{Hansch1999,Hansch2001,Hansch2012}.

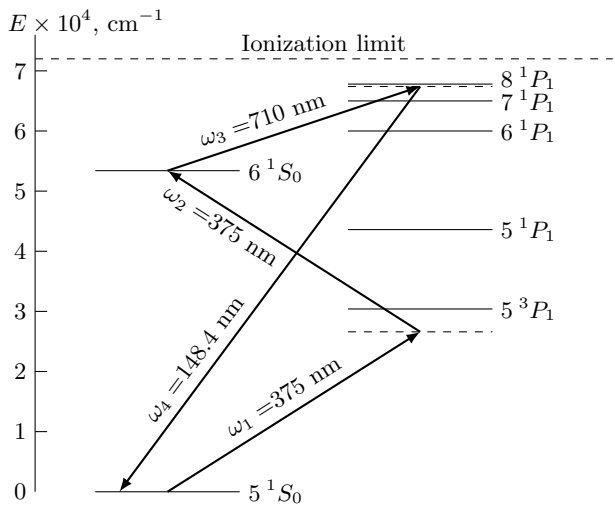
\begin{figure}

    \begin{tikzpicture}[scale=1.6]

 \draw (0,0)--(0,3.8);
\foreach \y in {0.5,1,...,3.5}
  \draw (0,\y)--(0.1,\y);
   \node [anchor=east] at (-0.0,0) {0};
   \node [anchor=east] at (-0.0,0.5) {1};
   \node [anchor=east] at (-0.0,1) {2};
   \node [anchor=east] at (-0.0,1.5) {3};
 \node [anchor=east] at (-0.0,2) {4};
 \node [anchor=east] at (-0.0,2.5) {5};
  \node [anchor=east] at (-0.0,3) {6};
  \node [anchor=east] at (-0.0,3.5) {7};
  \node [anchor=west] at (-0.3,3.9) {$E \times 10^{4}$, cm$^{-1}$};
  \draw (0.5,0) -- (1.7, 0) node at (2.0,0) {5$\:^1 S_0$};
  \draw [dashed] (0,3.6) -- (4.8,3.6);
  \draw (2.6,1.52) -- (3.8, 1.52) node at (4.1,1.52) {5$\:^3 P_1$};
  \draw (2.6,2.18) -- (3.8, 2.18) node at (4.1,2.18) {5$\:^1 P_1$};
  \draw (2.6,3) -- (3.8, 3) node at (4.1,3) {6$\:^1 P_1$};
  \draw (2.6,3.25) -- (3.8, 3.25) node at (4.1,3.25) {7$\:^1 P_1$};
  \draw (2.6,3.39) -- (3.8, 3.39) node at (4.1,3.44) {8$\:^1 P_1$};
  \draw [dashed] (2.6,3.37) -- (3.8, 3.37);
  \draw [dashed] (2.6,1.33) -- (3.8, 1.33);

   \draw (0.5,2.67) -- (1.7, 2.67) node at (2.0,2.67) {6$\:^1 S_0$};

  \draw [-latex, thick] (1.1,0) -- (3.2,1.33) node [midway, above, sloped, color=black] (TextNode) {$\omega_1=$375 nm};
  \draw [-latex, thick] (3.2,1.33) -- (1.1,2.67) node [pos=0.75, below, sloped, color=black] (TextNode) {$\omega_2=$375 nm};
  \draw [-latex, thick] (1.1,2.67) -- (3.2,3.37) node [pos=0.4, above, sloped, color=black] (TextNode) {$\omega_3=$710 nm};
  \draw [-latex, thick] (3.2,3.37) -- (0.7,0) node [pos=0.7, above, sloped, color=black] (TextNode) {$\omega_4=$148.4 nm};
  \node at (2.4,3.73) {Ionization limit};
\end{tikzpicture}
   \caption{Simplified energy level scheme of neutral Cd. The arrows represent incident lasers with frequencies $\omega_1$, $\omega_2$, $\omega_3$ and the generated VUV laser with frequency $\omega_4$ at 148.4 nm. }
    \centering
    \label{Scheme_6s}
\end{figure}

For the FWM process considered here, a new wave with the sum frequency of three incident laser beams is generated, $\omega_{1}+\omega_{2}+\omega_{3} \rightarrow \omega_{4}$, as shown in Fig.~\ref{Scheme_6s}.
For tightly focused incident laser beams with equal confocal parameter $b$, the FWM output power in the small signal limit~\cite{VidalReview} is given by~\cite{Bjorklund}
\begin{equation}
    P_{4}=\frac{9}{4}\frac{\omega_{1}\omega_{2}\omega_{3}\omega_{4}}{\pi^{2}\epsilon_{0}^{2}c^{6}}\frac{1}{b^{2}}\biggl(\frac{1}{\Delta k_{a}}\biggr)^{2}|\chi_{a}^{(3)}|^{2}P_{1}P_{2}P_{3}G(bN\Delta k_{a})\label{FWMoutput},
\end{equation}
where $\omega_{i}$ is the laser angular frequency, $P_{i}$ the laser power, $N$ the number density of the nonlinear medium, $\Delta k_{a}=(k_{4}-k_{1}-k_{2}-k_{3})/N$ the wave vector mismatch per atom, $\chi_{a}^{(3)}$ the third-order nonlinear susceptibility per atom, and $G(bN\Delta k_a)$ the phase matching function.

To understand the FWM process described by Eq.~\ref{FWMoutput}, the phase matching function $G(bN\Delta k_a)$ and the calculation of nonlinear susceptibility $\chi_{a}^{(3)}$ are detailed in the following.
The wave vector mismatch $\Delta k_{a}$ due to dispersion indicates the momentum mismatch in nonlinear optical processes. In the tight focus limit, where the confocal length of the laser beams $b$ is much smaller than the length of the nonlinear medium $L$, this mismatch can be compensated by the Gouy phase shift.
In this limit, the Gouy phase shift is experienced three times by the incident beams in the Gaussian focus but only once by the frequency-summed generated beam~\cite{Hansch2005}. Thus, the phase matching function $G(bN\Delta k_a)$ is maximized not at $\Delta k_{a} = 0$ as in the nonlinear optical processes with the collimated beams, but at $b N \Delta k_{a} = -4$.
It is important to note that $\Delta k_{a}$ does not depend on the number density of the nonlinear medium (see Supplemental Material~\cite{SupplMat}) but the frequencies of incident and generated lasers. 
Therefore, the optimal phase matching condition can be achieved by adjusting $b$ and $N$ accordingly~\cite{Bjorklund}.

The calculation of the nonlinear susceptibility $\chi_{a}^{(3)}$ from the first principles can be found in~\cite{BoydNLO}. We employ the two-photon resonance condition, $\omega_1+\omega_2=\omega_{rg}$, to enhance the output VUV power~\cite{TwoResonance}, with $6\: ^{1}S_{0}$ as the resonance state (denoted as $r$, with  $\omega_{rg}$ representing its transition frequency to the ground state $g$).
For three linearly polarized incident laser beams, the leading term in $\chi_{a}^{(3)}$ is~\cite{AVSmith1987}
\begin{equation}
    \chi_a^{(3)}=\frac{1}{6 \epsilon_0\hbar^3}S(\omega_1+\omega_2)\chi_{12}\chi_{34},
    \label{chi}
\end{equation}
with
\begin{equation}
    S(\omega_1+\omega_2)=\frac{1}{\Omega_{rg}-(\omega_1 + \omega_2)},
    \label{S_shape_equation}
\end{equation}
\begin{equation}
    \chi_{12}=\sum_l\left(\frac{\mu_{rl}\mu_{lg}}{\omega_{lg}-\omega_1}+\frac{\mu_{rl}\mu_{lg}}{\omega_{lg}-\omega_2}\right),
    \label{Chi12}
\end{equation}
\begin{equation}
    \chi_{34}=\sum_{m}\left(\frac{\mu_{rm}\mu_{mg}}{\omega_{mg}-\omega_4}+\frac{\mu_{rm}\mu_{mg}}{\omega_{mg}+\omega_3}\right).
    \label{Chi34}
\end{equation}
Here, $\Omega_{rg}=\omega_{rg}-\mathrm{i}\Gamma_{r}/2$ is the complex transition frequency containing the transition frequency $\omega_{rg}$ and the total population decay rate of the $r$ state $\Gamma_{r}$, and $\mu_{ij}$ are the $z$ components of the transition matrix elements coupling states $i$ and $j$.
The summation is performed over all the intermediate states $l(m)$ (namely, states $n \: ^{1,3} P_1$), whose transition frequencies to the ground state $g$ are denoted as $\omega_{lg}(\omega_{mg})$.

$S(\omega_1 + \omega_2)$ describes the shape of the two-photon resonance. In Eq.~\ref{S_shape_equation}, we do not consider pressure broadening and Doppler broadening for simplicity. 
The pressure broadening $\Delta\omega_\mathrm{p}$ 
can be included in the expression of the imaginary part of $\Omega_{rg}$ by $\Gamma^{'}_r =\Gamma_r+2\Delta\omega_\mathrm{p}$,
%and dictates the linewidth of $S(\omega_1 + \omega_2)$, 
while the Doppler broadening, which contributes through the Doppler shift of $\omega_1$ and $\omega_2$, can be introduced by a convolution of $S(\omega_1 + \omega_2)$ with the Maxwell-Boltzmann velocity distribution function 
% Maybe we should reconsider if 'decay rate' is the proper word
(see Supplemental Material~\cite{SupplMat} for a full mathematical treatment), and dominates the linewidth of $S(\omega_1 + \omega_2)$. 
Both broadening effects reduce $S(\omega_1+\omega_2)$, and, consequently, the FWM conversion efficiency. 
% Both broadening effects reduce $S(\omega_1+\omega_2)$, and so does the FWM conversion efficiency.
For cadmium vapor with natural abundance, since the isotope shifts are typically larger than or comparable to the Doppler broadening linewidth, $S(\omega_1+\omega_2)$ should be calculated for each isotope and weighted according to their relative abundances in the summation.  
To our knowledge, isotope shifts of the $5\: ^{1}S_{0} - 6\: ^{1}S_{0}$ transition in cadmium have not been measured.
In our computations, we assume the use of pure $^{114} \mathrm{Cd}$ isotope as the nonlinear medium, which is commercially available at a reasonable price. 
This should result in higher output VUV power than in the natural abundance case~\cite{Hansch2005}.

For experimental simplicity, we assume $\omega_1=\omega_2$, where only two incident laser beams at 375 nm and 710 nm are required to produce the VUV light (see Fig.~\ref{Scheme_6s}). For the more general case where $\omega_1 \neq \omega_2$, see Supplemental Material~\cite{SupplMat}.

Theoretically, the summation in Eqs.~\ref{Chi12} and~\ref{Chi34} is carried out over an infinite number of intermediate states. In our calculation, however, the summation is done over 20 intermediate $P$-states, namely $n\: ^{1}P_{1}$ and $n\: ^{3}P_{1}$ states with $n = 5 - 14$, since the transition matrix elements for high-lying states decrease substantially and the transition frequencies of the corresponding states are detuned from the generated VUV light.
Considering the dipole approximation and assuming the linear laser polarization, the selection rules for the coupled states are $\Delta J = \pm 1$ and $\Delta M_J = 0$. All corresponding matrix elements (see Supplemental Material~\cite{SupplMat}) are calculated using the relativistic Fock-space coupled cluster (FS-CC) method implemented in {\sc dirac}~\cite{DIRAC19,Saue:2020} and Exp-T~\cite{Oleynichenko_EXPT,EXPT_website} program packages. 
A more detailed investigation of relativistic and electron correlation effects for high-lying states will be published elsewhere.

The relative contributions to $\chi_{12}$ and $\chi_{34}$ from various intermediate $P$-states are shown in Fig.~\ref{ablation_study_6s}. 
The dominant contributions to $\chi_{12}$ and $\chi_{34}$ originate from the $5\: ^1 P_1$ and $6\: ^1 P_1$ states. For $\chi_{34}$, the $7\: ^1 P_1$ and $8\: ^1 P_1$ states each contribute approximately 10$\%$.
Contributions from all triplet states and states above the $8\: ^1 P_1$ to both $\chi_{12}$ and $\chi_{34}$ are minor.

\begin{figure}
    \centering
       \begin{tikzpicture}
\begin{axis}[symbolic x coords={$5\: ^{3} P_1$,$5\: ^{1} P_1$,$6\: ^{3} P_1$,$6\: ^{1} P_1$,$7\: ^{3} P_1$,$7\: ^{1} P_1$,$8\: ^{3} P_1$,$8\: ^{1} P_1$,$9\: ^{3} P_1$,$9\: ^{1} P_1$,$10\: ^{3} P_1$,$10\: ^{1} P_1$},
xtick={$5\: ^{3} P_1$,$5\: ^{1} P_1$,$6\: ^{3} P_1$,$6\: ^{1} P_1$,$7\: ^{3} P_1$,$7\: ^{1} P_1$,$8\: ^{3} P_1$,$8\: ^{1} P_1$,$9\: ^{3} P_1$,$9\: ^{1} P_1$,$10\: ^{3} P_1$,$10\: ^{1} P_1$}, 
xticklabel style={align=center,rotate=90},
ymin=0,
ytick={0,0.1,...,0.8},
ylabel style={inner ysep=5pt},
ylabel near ticks,
ylabel=Relative contribution,
bar width=5pt,
width=0.5\textwidth] 

\addplot[ybar,fill=blue,bar shift=-2.5pt] coordinates {
($5\: ^{3} P_1$,0.00289)
($5\: ^{1} P_1$,0.810488)
($6\: ^{3} P_1$,0.00026)
($6\: ^{1} P_1$,0.174)
($7\: ^{3} P_1$,0.0000006)
($7\: ^{1} P_1$,0.00939)
($8\: ^{3} P_1$,0.000000006)
($8\: ^{1} P_1$,0.00198)
($9\: ^{3} P_1$,0.00000007)
($9\: ^{1} P_1$,0.00071)
($10\: ^{3} P_1$,0.00000005)
($10\: ^{1} P_1$,0.000362)

};
\addplot[ybar,fill=green,bar shift=2.5pt] coordinates {
($5\: ^{3} P_1$,0.000284)
($5\: ^{1} P_1$,0.4074)
($6\: ^{3} P_1$,0.0005)
($6\: ^{1} P_1$,0.422)
($7\: ^{3} P_1$,0.000005)
($7\: ^{1} P_1$,0.098)
($8\: ^{3} P_1$,0.0000003)
($8\: ^{1} P_1$,0.0617)
($9\: ^{3} P_1$,0.0000007)
($9\: ^{1} P_1$,0.0076)
($10\: ^{3} P_1$,0.0000004)
($10\: ^{1} P_1$,0.0028)
};
\end{axis}
\end{tikzpicture}
\caption{The normalized relative contributions to $\chi_{12}$ (blue bars) and $\chi_{34}$ (green bars) from a set of intermediate $P$-states up to the $10 ^{1} P_1$. 
}
\label{ablation_study_6s}
\end{figure}
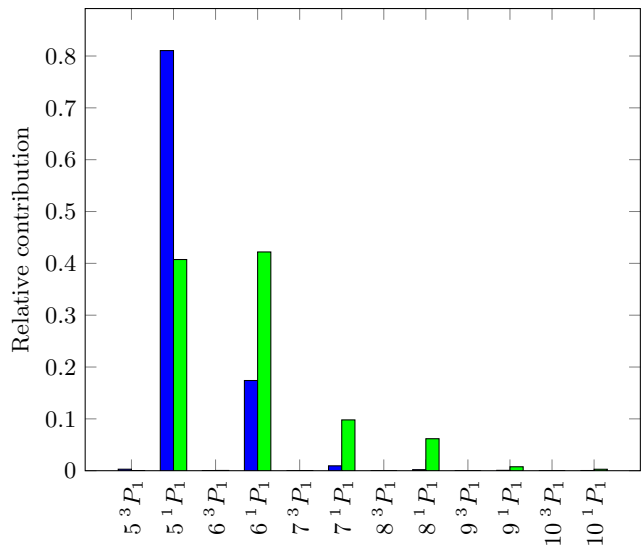

We choose an incident laser power of 3~W at 375~nm and 6~W at 710~nm,  which are readily available in the laboratory. 
The confocal parameter is set to $b=1.5$ mm, and the vapor temperature is set to $T = 580$~$^\circ\mathrm{C}$ to optimize the output VUV power. The saturation vapor pressure as a function of temperature can be found in the Supplemental Material~\cite{SupplMat}.
Under these conditions, we calculate the nonlinear susceptibility $\chi_{a}^{(3)}$, the phase matching function $G(bN\Delta k_a)$, and the generated VUV power $P_{4}$ in the vicinity of Th-229 isomer transition, as shown in Fig.~\ref{powers}.

The nonlinear susceptibility $\chi_{a}^{(3)}$ exhibits divergence near resonant states since the absorption is not accounted for in Eq.~\ref{chi}.
The phase-matching function $G(bN\Delta k_a)$ goes to zero near resonant states because the refractive index of the generated beam is so large that the resulting wave vector mismatch $\Delta k_a$ cannot be compensated for using the current parameters. 
The calculated output power $P_{4}$ at 148.4~nm is approximately 30 $\mathrm{\mu W}$. 
Zero VUV power points arise when terms with different signs cancel in the $\chi_{34}$ expression shown in Eq.~\ref{Chi34}.

The VUV power can be further enhanced by introducing some noble gas to provide additional compensation for wave vector mismatch, or by tuning one of the incident beams close to the $5 \:^{1,3}P_1$ resonance states, while maintaining two-photon resonance with the second incident beam. For a detailed discussion, see Supplemental Material~\cite{SupplMat}.

We also perform calculations for the FWM scheme exploiting $5\: ^1 D_2$ as the two-photon resonance state, which is described in detail in the Supplemental Material~\cite{SupplMat}.
Similar to the case with $6\: ^{1}S_{0}$ as the resonance state, 
the coupling to $5\: ^1 P_1$ and $6\: ^1 P_1$ states dominates the nonlinear susceptibility, and the output VUV power is at comparable level.  

\begin{figure}
    \centering
    \includegraphics{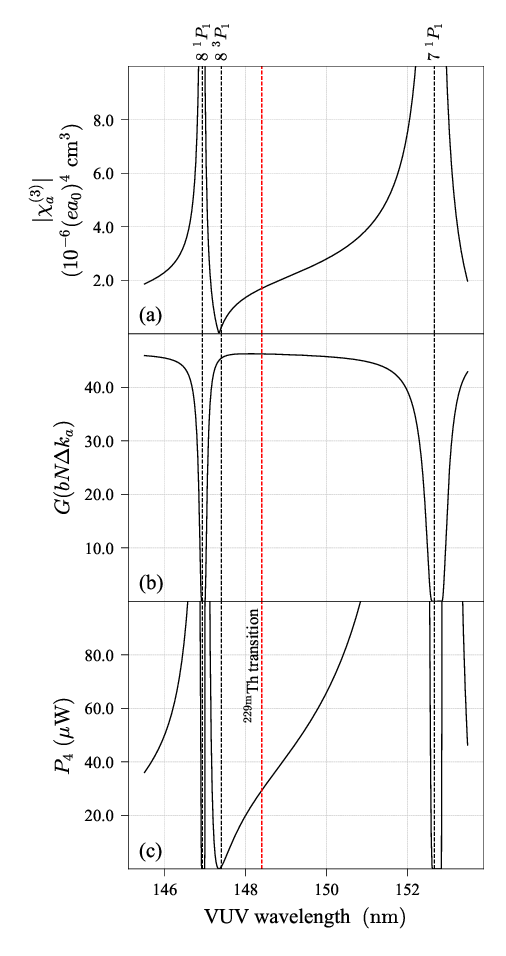}
     \caption{Calculated nonlinear susceptibility $\left| \chi_{a}^{(3)} \right|$ , phase matching function $G(bN\Delta k_a)$, and VUV yield $P_{4}$ in the vicinity of Th-229 isomer transition according to Eq.~\ref{FWMoutput} using parameters stated in the text. Note that the effects of absorption, saturation are ignored in the plot.}
    \label{powers}
\end{figure}

The output VUV power in Fig.~\ref{powers}(c) does not account for absorption and saturation effects. 
Saturation might be an issue in the tightly focused case, where the laser intensity near the beam waist can be enormous. 
We find that the only significant saturation effect under the current experimental parameters is due to the depletion of the ground state induced by the two-photon resonance Raman-type absorption~\cite{SupplMat}. 
This issue can be mitigated either by slightly detuning the lasers from the exact two-photon resonance condition, or by loosening the focus while adding some noble gas into the medium for phase mismatch compensation.
Other effects, such as the linear absorption of the beams, same order and higher order nonlinear processes, and amplified spontaneous emission are found to be insignificant~\cite{SupplMat}.

As a conservative estimate, we assume a VUV power of 10~$\mu$W focused to a spot diameter of $8$~$\mu$m to study the interaction of the Th-229 nucleus with the VUV light. This interaction has been treated in detail in~\cite{LarsCalculation}.
We consider the Th-229 isomer transition as the simplest two-level system interacting with a VUV laser with intensity $I_l$ and linewidth $\Gamma_l$. In order to observe nuclear Rabi oscillations, the linewidth of the VUV laser $\Gamma_{l}$ needs to satisfy
\begin{equation}
    \Gamma_{l} < 2\Omega_{eg} = 2\sqrt{\frac{2 \pi c^2 I_{l} \Gamma_\gamma}{\hbar \omega_0^3}} \approx 1\,\text{kHz},
    \label{linewidthcondition}
\end{equation}
where $\Omega_{eg}$ is the Rabi frequency, $\Gamma_{\gamma}$ the radiative decay rate of the nuclear excited state, and $\omega_0$ the angular frequency of the isomer transition.
The nuclear excitation probability as a function of VUV irradiation time for laser linewidths below and above 1 kHz can be found in the Supplemental Material~\cite{SupplMat}.

We argue that the linewidth condition Eq.~\ref{linewidthcondition} can be satisfied in the FWM process. 
Previously, it has been extensively studied and shown~\cite{HHG_cutoff,PhaseHHG,DipolePhase2006,Mech_Below,Below_threshold_noble,Below_threshold_dynamics} that HHG processes in gases are governed by the coherent temporal dynamics of the electron. 
Advances in VUV frequency comb technology~\cite{VUVYe2005,Yost_VUV_HHG,VUVCoherence2014} have demonstrated the capability for generating VUV radiation with coherence time greater than 1~s, allowing for sub-hertz spectral resolution for HHG processes.
Particularly interesting for the current work is the third harmonic generation~\cite{VUVYe2005}, which also relies on the third-order nonlinearity, and is a special case of FWM process when $\omega_1=\omega_2=\omega_3$.
Similar to other noiseless classical frequency multiplications, the phase noise of the incident beams will be scaled up in the FWM process~\cite{PeikFWM}, but this incident phase noise can be substantially suppressed by locking the lasers to ultra-stable high-finesse cavities.
Therefore, we expect the FWM process to be highly coherent, resulting in VUV light with a linewidth below 1 kHz.

In conclusion, we propose using four-wave mixing in cadmium vapor to generate continuous-wave laser light for precision spectroscopy of the Th-229 isomer transition.
Near the Th-229 isomer transition frequency, the nonlinear susceptibility of the employed FWM scheme is large, and the phase-matching condition can be satisfied.
With an incident laser power of 3 W at 375 nm and 6 W at 710 nm, more than 30 $\mu$W of VUV light at 148.4 nm can be produced.
The resulting VUV power can be further enhanced by coupling the incident laser beams into a high-finesse cavity to increase the effective incident laser power. 
The generated VUV light with a sub-kHz linewidth will enable precision spectroscopy of Th-229 isomer transition, the driving of nuclear Rabi oscillations in an ion trap~\cite{Th3JapanNature,KuzmichCrystal,IonPeik}, the observation of the collective effects of nuclear transitions in thorium-doped crystals~\cite{collective} and the construction of a nuclear optical clock~\cite{ClockPeik,ClockKuzmich}. 
Additionally, the calculation method can be applied to other nonlinear media to produce VUV light for various applications, such as laser cooling of Al$^+$ ions and high-resolution atomic and molecular spectroscopy.

While finishing the current work, we became aware of Jun Ye's group's research on Th-229 isomer spectroscopy using a VUV frequency comb, as reported at DAMOP 2024~\cite{Ooi2024APS}.

\begin{acknowledgments}
We thank Peter Thirolf, Jun Ye, Leonid Skripnikov, Li You and Haoyu Shi for helpful discussions, and Peter Thirolf for carefully reading the manuscript. This work is supported by the National Natural Science Foundation of China (Grants No. 12341401 and No. 12274253) and the National Key Research and Development Program of China (2021YFA1402104). 
\end{acknowledgments}

\clearpage

\title{Supplemental material}

\maketitle

\begin{table*}
\caption{Values of reduced E1 matrix elements for neutral Cd used in the calculation of the nonlinear susceptibility.}
\newcolumntype{Y}{>{\centering\arraybackslash}X}
\begin{tabularx}{0.9\textwidth}{c *{6}{Y}}
\hline
\hline
 Transition & d, a.u. & Transition & d, a.u. & Transition & d, a.u. \\
\hline
$5s\: ^{1}S_0 \to 5\: ^{3}P_1$ & 0.148 & $5\: ^{3}P_1 \to  6\: ^{1}S_0$ & 0.082 & $5\: ^{3}P_1 \to 5\: ^{1}D_2$ & 0.138 \\
$5\: ^{1}S_0 \to 5\: ^{1}P_1$ & 3.461 & $5\: ^{1}P_1 \to 6\: ^{1}S_0$ & 4.199 & $5\: ^{1}P_1 \to 5\: ^{1}D_2$ & 5.673 \\
$5\: ^{1}S_0 \to 6\: ^{3}P_1$ & 0.013 & $6\: ^{1}S_0 \to 6\: ^{3}P_1$ & 0.696 & $6\: ^{3}P_1 \to 5\: ^{1}D_2$ & 0.111 \\
$5\: ^{1}S_0 \to 6\: ^{1}P_1$ & 0.769 & $6\: ^{1}S_0 \to 6\: ^{1}P_1$ & 7.919 & $5\: ^{1}D_2 \to 6\: ^{1}P_1$ & 12.241 \\
$5\: ^{1}S_0 \to 7\: ^{3}P_1$ & 0.001 & $6\: ^{1}S_0 \to 7\: ^{3}P_1$ & 0.021 & $5\: ^{1}D_2 \to 7\: ^{3}P_1$ & 0.069 \\
$5\: ^{1}S_0 \to 7\: ^{1}P_1$ & 0.327 & $6\: ^{1}S_0 \to 7\: ^{1}P_1$ & 1.175 & $5\: ^{1}D_2 \to 7\: ^{1}P_1$ & 0.552 \\
$5\: ^{1}S_0 \to 8\: ^{3}P_1$ & 0.0004 & $6\: ^{1}S_0 \to 8\: ^{3}P_1$ & 0.007 & $5\: ^{1}D_2 \to 8\: ^{3}P_1$ & 0.034 \\
$5\: ^{1}S_0 \to 8\: ^{1}P_1$ & 0.196 & $6\: ^{1}S_0 \to 8\: ^{1}P_1$ & 0.441 & $5\: ^{1}D_2 \to 8\: ^{1}P_1$ & 0.149 \\
$5\: ^{1}S_0 \to 9\: ^{3}P_1$ & 0.0003 & $6\: ^{1}S_0 \to 9\: ^{3}P_1$ & 0.010 & $5\: ^{1}D_2 \to 9\: ^{3}P_1$ & 0.020 \\
$5\: ^{1}S_0 \to 9\: ^{1}P_1$ & 0.136 & $6\: ^{1}S_0 \to 9\: ^{1}P_1$ & 0.237 & $5\: ^{1}D_2 \to 9\: ^{1}P_1$ & 0.27 \\
$5\: ^{1}S_0 \to 10\: ^{3}P_1$ & 0.0003 & $6\: ^{1}S_0 \to 10\: ^{3}P_1$ & 0.009 & $5\: ^{1}D_2 \to 10\: ^{3}P_1$ & 0.003 \\
$5\: ^{1}S_0 \to 10\: ^{1}P_1$ & 0.106  & $6\: ^{1}S_0 \to 10\: ^{1}P_1$ & 0.157  & $5\: ^{1}D_2 \to 10\: ^{1}P_1$ & 0.099 \\
$5\: ^{1}S_0 \to 11\: ^{3}P_1$ & 0.0002  & $6\: ^{1}S_0 \to 11\: ^{3}P_1$ & 0.006  & $5\: ^{1}D_2 \to 11\: ^{3}P_1$ & 0.007 \\
$5\: ^{1}S_0 \to 11\: ^{1}P_1$ & 0.081  & $6\: ^{1}S_0 \to 11\: ^{1}P_1$ & 0.112  & $5\: ^{1}D_2 \to 11\: ^{1}P_1$ & 0.038 \\
$5\: ^{1}S_0 \to 12\: ^{3}P_1$ & 0.0003  & $6\: ^{1}S_0 \to 12\: ^{3}P_1$ & 0.003  & $5\: ^{1}D_2 \to 12\: ^{3}P_1$ & 0.005 \\
$5\: ^{1}S_0 \to 12\: ^{1}P_1$ & 0.050  & $6\: ^{1}S_0 \to 12\: ^{1}P_1$ & 0.070  & $5\: ^{1}D_2 \to 12\: ^{1}P_1$ & 0.012 \\
$5\: ^{1}S_0 \to 13\: ^{3}P_1$ & 0.0005  & $6\: ^{1}S_0 \to 13\: ^{3}P_1$ & 0.001  & $5\: ^{1}D_2 \to 13\: ^{3}P_1$ & 0.003 \\
$5\: ^{1}S_0 \to 13\: ^{1}P_1$ & 0.028  & $6\: ^{1}S_0 \to 13\: ^{1}P_1$ & 0.041  & $5\: ^{1}D_2 \to 13\: ^{1}P_1$ & 0.003 \\
$5\: ^{1}S_0 \to 14\: ^{3}P_1$ & 0.0005 & $6\: ^{1}S_0 \to 14\: ^{3}P_1$ & 0.0006 &  $5\: ^{1}D_2 \to 14\: ^{3}P_1$ & 0.002 \\
$5\: ^{1}S_0 \to 14\: ^{1}P_1$ & 0.017 &  $6\: ^{1}S_0 \to 14\: ^{1}P_1$ & 0.027 &  $5\: ^{1}D_2 \to 14\: ^{1}P_1$ & 0.001 \\
\hline
\hline
\end{tabularx}
\label{table:3}
\end{table*}

\section*{Supplemental material}
\subsection{$5\: ^{1}D_2$ as the two-photon resonance state}\label{5D_channel}
We have also performed calculations for the case with $5\: ^{1}D_{2}$ as the two-photon resonance state. 
For $\omega_1=\omega_2$, the incident lasers are $338 ~\mathrm{nm}$ and $1223 ~\mathrm{nm}$, as shown in Fig.~\ref{Scheme}. 

\begin{figure}
    \centering
       \begin{tikzpicture}
\begin{axis}[symbolic x coords={$5\: ^{3} P_1$,$5\: ^{1} P_1$,$6\: ^{3} P_1$,$6\: ^{1} P_1$,$7\: ^{3} P_1$,$7\: ^{1} P_1$,$8\: ^{3} P_1$,$8\: ^{1} P_1$,$9\: ^{3} P_1$,$9\: ^{1} P_1$,$10\: ^{3} P_1$,$10\: ^{1} P_1$},
xtick={$5\: ^{3} P_1$,$5\: ^{1} P_1$,$6\: ^{3} P_1$,$6\: ^{1} P_1$,$7\: ^{3} P_1$,$7\: ^{1} P_1$,$8\: ^{3} P_1$,$8\: ^{1} P_1$,$9\: ^{3} P_1$,$9\: ^{1} P_1$,$10\: ^{3} P_1$,$10\: ^{1} P_1$}, 
xticklabel style={align=center,rotate=90},
ymin=0,
ytick={0,0.1,...,0.8},
ylabel style={inner ysep=5pt},
ylabel near ticks,
ylabel=Relative contribution,
bar width=5pt,
width=0.5\textwidth] 

\addplot[ybar,fill=blue,bar shift=-2.5pt] coordinates {
($5\: ^{3} P_1$,0.011)
($5\: ^{1} P_1$,0.8053)
($6\: ^{3} P_1$,0.00003)
($6\: ^{1} P_1$,0.1796)
($7\: ^{3} P_1$,0.000002)
($7\: ^{1} P_1$,0.0029)
($8\: ^{3} P_1$,0.00000009)
($8\: ^{1} P_1$,0.00044)
($9\: ^{3} P_1$,0.00000009)
($9\: ^{1} P_1$,0.00054)
($10\: ^{3} P_1$,0.00000001)
($10\: ^{1} P_1$,0.00015)
};

\addplot[ybar,fill=green,bar shift=2.5pt] coordinates {
($5\: ^{3} P_1$,0.00039)
($5\: ^{1} P_1$,0.436)
($6\: ^{3} P_1$,0.00006)
($6\: ^{1} P_1$,0.504)
($7\: ^{3} P_1$,0.00001)
($7\: ^{1} P_1$,0.035)
($8\: ^{3} P_1$,0.000001)
($8\: ^{1} P_1$,0.016)
($9\: ^{3} P_1$,0.000001)
($9\: ^{1} P_1$,0.0068)
($10\: ^{3} P_1$,0.0000001)
($10\: ^{1} P_1$,0.0014)
};
\end{axis}
\end{tikzpicture}
\caption{The normalized relative contributions to $\chi_{12}$ (blue bars) and $\chi_{34}$ (green bars) for the scheme exploiting the $5\: ^{1}D_{2}$ as the two-photon resonance state from a set of intermediate $P$-states up to $10\:^{1} P_1$. 
}
\label{ablation_study_5d}
\end{figure}
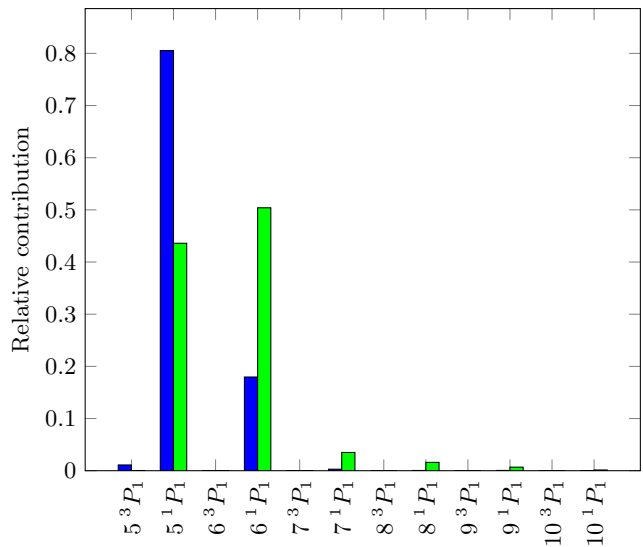

The relative contributions to $\chi_{12}$ and $\chi_{34}$ stemming from different intermediate $P$-states are presented in Fig.~\ref{ablation_study_5d}. 
Similar to the case with $6\: ^{1}S_{0}$ as the two-photon resonance state, the coupling to the $5\: ^1 P_1$ and $6\: ^1 P_1$ states dominates the nonlinear susceptibility, and contributions from all triplet states and states above $8\: ^1 P_1$ are negligible.

\begin{figure}[H]
    \centering
    \begin{tikzpicture}[scale=1.6]

 \draw (0,0)--(0,5.7);
\foreach \y in {0.75,1.5,...,5.25}
  \draw (0,\y)--(0.1,\y);
   \node [anchor=east] at (-0.0,0) {0};
   \node [anchor=east] at (-0.0,0.75) {1};
   \node [anchor=east] at (-0.0,1.5) {2};
   \node [anchor=east] at (-0.0,2.25) {3};
 \node [anchor=east] at (-0.0,3) {4};
 \node [anchor=east] at (-0.0,3.75) {5};
  \node [anchor=east] at (-0.0,4.5) {6};
  \node [anchor=east] at (-0.0,5.25) {7};
  \node [anchor=west] at (-0.3,5.9) {$E \times 10^{4}$, cm$^{-1}$};
  \draw (0.4,0) -- (1.2, 0) node at (1.5,0) {5$\:^1 S_0$};
  \draw [dashed] (0,5.4) -- (4.8,5.4);
  \draw (2.0,2.28) -- (2.8, 2.28) node at (3.1,2.28) {5$\:^3 P_1$};
  \draw (2.0,3.27) -- (2.8, 3.27) node at (3.1,3.27) {5$\:^1 P_1$};
  \draw (2.0,4.49) -- (2.8, 4.49) node at (3.1,4.49) {6$\:^1 P_1$};
  \draw (2.0,4.875) -- (2.8, 4.875) node at (3.1,4.875) {7$\:^1 P_1$};
  \draw (2.0,5.09) -- (2.8, 5.09) node at (3.1,5.09) {8$\:^1 P_1$};
  \draw [dashed] (2.0,5.055) -- (2.8, 5.055);
  \draw [dashed] (2.0,2) -- (2.8, 2);
  \draw [dashed] (2.0,2.22) -- (2.8, 2.22);

   \draw (0.4,4) -- (1.2, 4) node at (1.5,4) {6$\:^1 S_0$};
   \draw (3.6,4.44) -- (4.4,4.44) node at (4.7,4.44) {5$\:^1 D_2$};

  \draw [-latex, thick, red] (0.7,0) -- (2.3,2) node [midway, above, sloped, color=black] (TextNode) {$\omega_1=$375 nm};
  \draw [-latex, thick, red] (2.3,2) -- (0.6,4) node [pos=0.65, below, sloped, color=black] (TextNode) {$\omega_2=$375 nm};
  \draw [-latex, thick, red] (0.6,4) -- (2.4,5.055) node [midway, above, sloped, color=black] (TextNode) {$\omega_3=$710 nm};
  \draw [-latex, thick, blue] (0.9,0) -- (2.674,2.22) node [midway, below, sloped, color=black] (TextNode) {$\omega_1=$338 nm};
  \draw [-latex, thick, blue] (2.674,2.22) -- (4.2,4.44) node [midway, below, sloped, color=black] (TextNode) {$\omega_2=$338 nm};
  \draw [-latex, thick, blue] (4.2,4.44) -- (2.4,5.055) node [pos=0.2, above, sloped, color=black] (TextNode) {$\omega_3=$1223 nm};
  \draw [-latex, thick] (2.4,5.055) -- (0.5,0) node [pos=0.7, above, sloped, color=black] (TextNode) {$\omega_4=$148.4 nm};
  \node at (2.4,5.6) {Ionization limit};
\end{tikzpicture}
   \caption{Simplified energy level scheme of neutral Cd. The red and blue arrows represent the incident lasers $\omega_1$, $\omega_2$, $\omega_3$ with $6\: ^{1}S_{0}$ and $5\: ^{1}D_{2}$ as two-photon resonance states, respectively. The black arrow indicates the generated VUV laser at 148.4 nm. }
    \label{Scheme}
\end{figure}
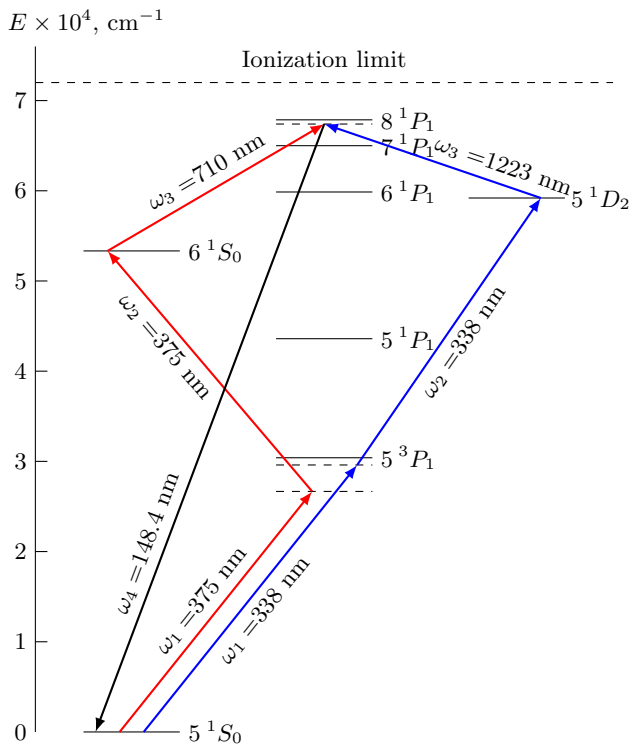

\begin{figure}
    \centering
    \includegraphics{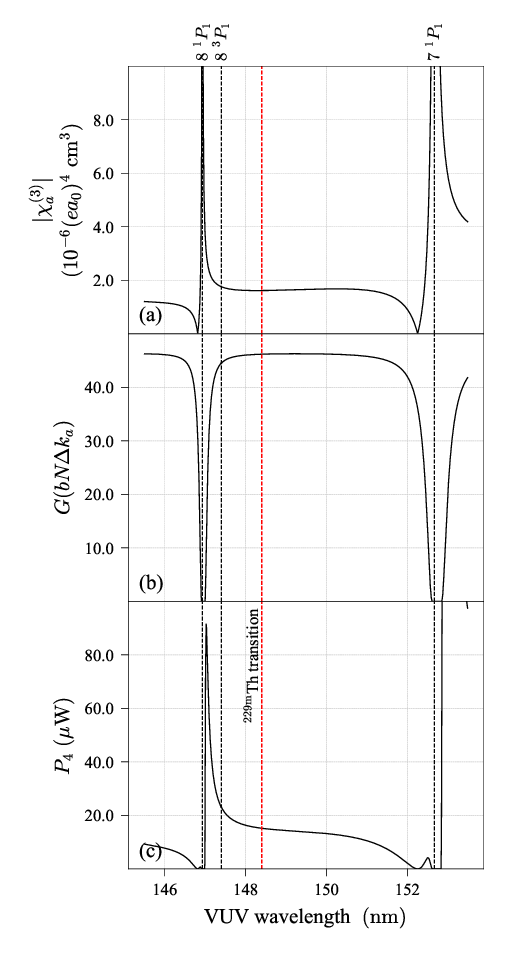}
    \caption{The calculated nonlinear susceptibility $\left| \chi_{a}^{(3)} \right|$ , the phase matching function $G(b N \Delta k_a)$, and the VUV yield $P_{4}$ in the vicinity of Th-229 isomer transition with the $5\: ^{1}D_{2}$ as the two-photon resonance state. The confocal parameter is $b = 1.5 ~\mathrm{mm}$, the temperature is $T = 570$~$\mathrm{^\circ C}$, and the power of the incident beams is $3 ~\mathrm{W}$ for the $338 ~\mathrm{nm}$ beam and $6 ~\mathrm{W}$ for the $1223 ~\mathrm{nm}$ beam.
    }
    \label{powerd}
\end{figure}

The calculated phase matching function $G(b N \Delta k_{a})$, the nonlinear susceptibility $\chi_{a}^{(3)}$, and the generated VUV power $P_{4}$ in the vicinity of Th-229 isomer transition are shown in Fig.~\ref{powerd}. The optimal phase matching temperature is $T = 570$~$\mathrm{^\circ C}$ for $b = 1.5~\mathrm{mm}$. 
With the same laser powers (3 W at 338 nm and 6 W at 1223 nm), around half of the VUV power can be generated compared to the case exploiting $6\: ^{1}S_{0}$ as the two-photon resonance state. 
Nevertheless, the power of the commercially available 1223 nm laser can be as high as 30 W, making the $5\: ^{1}D_2$ scheme also appealing.

\subsection{Phase matching}
It can be shown that the wave vector mismatch $\Delta k_{a}=(k_{4}-k_{1}-k_{2}-k_{3})/N$ is independent of the number density of the nonlinear medium $N$. 
The wave vector $k_{i}$ is related to the refractive index $n(\omega_{i})$ by
\begin{equation}
 k_{i}=n(\omega_{i})\omega_i/c,   
\end{equation}
where~\cite{Born_Principles_of_optics}
\begin{equation}
n(\omega_{i}) \simeq 1 + \frac{1}{2}\mathrm{Re}[\chi^{(1)}] = 1 + \frac{Ne^{2}}{2\epsilon_{0}m_{e}}\sum_{m}\frac{f_{m g}}{\omega_{m g}^{2}-\omega_{i}^{2}}.
\label{refractive_ind}
\end{equation}
Here $f_{m g}$ and $\omega_{m g}$ are the oscillator strength and transition frequency, respectively. 
% associated with the transition between states $m$ and $g$,
% is the corresponding transition frequency.
The oscillator strength $f_{m g}$ is given by~\cite{Born_Principles_of_optics}
\begin{equation}
    f_{mg}=\frac{2m\omega_{mg}|\boldsymbol{\mu}_{mg}|^2}{3\hbar e^2}, 
\end{equation}
with the dipole moment matrix element $\boldsymbol{\mu}_{mg}$ related to its $z$ component $\mu_{mg}$ by Wigner-Eckart Theorem~\cite{VidalReview}.

By making use of expression $\omega_{1}+\omega_{2}+\omega_{3}=\omega_{4}$, we get

\begin{equation}
\Delta k_{a} = \frac{e^{2}}{2\epsilon_{0}m_{e}c}\sum_{m} \biggl( \frac{f_{m g} \omega_{4}}{\omega_{m g}^{2}-\omega_{4}^{2}}-\sum_{j=1,2,3}\frac{f_{m g} \omega_{j}}{\omega_{m g}^{2}-\omega_{j}^{2}} \biggl) .
\end{equation}
Therefore, $\Delta k_{a}$ does not depend on the number density of the nonlinear medium, but the frequencies of incident and generated lasers. 

The phase-matching function in the tight focus limit can be written explicitly as~\cite{Bjorklund}
\begin{equation}
    \begin{aligned}&G(b N \Delta k_{a}) = G(b \Delta k)\\&=\begin{cases}0\quad\mathrm{for}\quad b\Delta k>0,\\\pi^2(b\Delta k)^4e^{b\Delta k}\quad\mathrm{for}\quad b\Delta k<0,\end{cases}\end{aligned}\label{Gfactor}
\end{equation}
where we define $\Delta k = k_{4}-k_{1}-k_{2}-k_{3} = N \Delta k_{a} $.
The function above attains its maximum value at $b\Delta k = -4$.

\subsection{Calculation of $S(\omega_1 + \omega_2)$}\label{SFactorSubsec}

The pressure broadening and Doppler broadening are typically much larger than the natural linewidth of the two-photon resonance state,
and must be considered for the calculation of $S(\omega_1+\omega_2)$.
The pressure broadening $\Delta\omega_\mathrm{p}$ is expressed as
\begin{equation}
    \Delta \omega_{\mathrm{p}}=N d^2 \sqrt{\frac{16 \pi k_b T}{m}},
\end{equation}
where $d$ and $m$ are the kinetic diameter and the atomic mass of the nonlinear medium atom, respectively, $T$ the medium temperature.
For the calculation of $S(\omega_1+\omega_2)$, pressure broadening can be directly added to the population decay rate $\Gamma_r $ by 
\begin{equation}
\Gamma^{'}_r =\Gamma_r+2\Delta\omega_\mathrm{p},
\label{PressBroad}
\end{equation}
since the decoherence of the collisional dephasing and the damping of the natural decay both contribute to decoherence of the transition. This results in a decrease of the FWM conversion effeciency.

The Doppler broadening is given by
\begin{equation}
    \Delta \omega_\mathrm{D} = 2\pi \omega_{rg} \sqrt{\frac{8 k_b T \ln{2}}{m c^2}}.
    \label{DoppBroad}
\end{equation}
It can be accounted for by using the Maxwell-Boltzmann distribution to collect contributions from atoms with different velocities~\cite{AVSmith1987}
\begin{equation}
    S(\omega_1+\omega_2)=\int_{-\infty}^{+\infty}\frac{1}{\Omega_{r g}-\omega_{1}'-\omega_{2}'}f(v_x)dv_x.\label{S_Doppler}
\end{equation}
Here $\omega_{i}'=\omega_{i}(1+\frac{v_x}{c})$ is the angular frequency in the atom frame moving with a longitudinal velocity $v_x$ towards the $i^{\mathrm{th}}$ incident beam, 
and $f(v_x)dv_x$ is the Maxwell-Boltzmann velocity distribution given by
\begin{equation}
    f(v_x)dv_x=\sqrt{\frac{m}{2\pi kT}}\exp\biggl(-\frac{mv_x^2}{2kT}\biggr)dv_{x}.
    \label{Maxwell Distribution}
\end{equation}

Substituting Eq.~\ref{PressBroad}, Eq.~\ref{DoppBroad} and Eq.~\ref{Maxwell Distribution} into Eq.~\ref{S_Doppler} and performing non-dimensionalization, we get~\cite{AVSmith1987}
\begin{equation}
    S(\omega_1+\omega_2)=\frac{1}{w}Z(\zeta).
    \label{shape_equation}
\end{equation}
 Here $Z(\zeta)$ is the plasma-dispersion function given by
\begin{equation}
    Z(\zeta)=\frac{1}{\sqrt{\pi}}\int_{-\infty}^{+\infty}dx\frac{e^{-x^{2}}}{x-\zeta},
\end{equation}
\begin{equation}
    \zeta=\bigg(\omega_1+\omega_2+i\frac{\Gamma_{r }^{'}}{2}-\omega_{r g}\bigg)/w,
\end{equation}
\begin{equation}
    w=\frac{\Delta \omega_{\mathrm{D}}}{2\sqrt{\ln2}}.
\end{equation}
Moving atoms experience detuning from the two-photon resonance condition, and contribute less to the value of $S(\omega_1+\omega_2)$. This leads to a Doppler-broadening-induced decrease of FWM conversion efficiency.

Note that $S(\omega_1 + \omega_2)$ has a weak correlation with the number density of the nonlinear medium $N$, since both the pressure broadening and the Doppler broadening depend on the medium temperature, which in turn dictates the number density of the nonlinear medium $N$.
Therefore, the optimal output power requires a phase-matching condition slightly deviating from the optimal phase-matching condition $b\Delta k = -4$ solely determined by optimizing $G(b N \Delta k_{a})$.

\subsection{Saturation and absorption effects}\label{Saturation_subsec}
In the calculation of the VUV power according to Eq.~1 in the main text, we assume the small signal limit, which is only valid when saturation and linear absorption effects are negligible. Saturation is defined as the deviation from the relation $P_4 \propto P_1 P_2 P_3$, which holds only when the intensities of the incident and generated beams are low.

At high incident laser intensities, two-photon resonance Raman-type absorption becomes significant, pumping the population from the $5\: ^1S_0$ state to the $6\: ^1S_0$ state. The depletion of the ground state reduces the third order nonlinear susceptibility $\chi_{a}^{(3)}$, thereby decreasing the output power.  
On the other hand, if the population in the $6\: ^1S_0$ state becomes sufficiently high to form a population inversion between the $6\: ^1S_0$ and the $5\: ^1P_1$ states, amplified spontaneous emission (ASE) may occur. ASE broadens the $6\: ^1S_0$ state, reducing the $S(\omega_1 + \omega_2)$ factor and the output power.
Additionally, the altered population distribution changes the refractive index $n$, potentially disrupting the phase-matching condition.

Other saturation mechanisms, such as absorption and refractive index changes induced by third-order and higher-order processes, as well as pump depletion arising from competing processes, are estimated to be negligible in our case~\cite{AVSmithOptimization}.

\subsubsection{Population introduced by two-photon absorption}\label{PopulationSubSec}

\begin{figure}
    \centering
    \begin{tikzpicture}[scale=1.2]

  \draw[thick] (0,0) -- (1,0) node[right] {$5\: ^1 S_0$};
  \draw[thick] (1.5,2) -- (2.5,2) node[right] {$5\: ^1 P_1$};
  \draw[thick] (0,4) -- (1,4) node[right] {$6\: ^1 S_0$};

  \draw[dashed] (1.5,2.3) -- (2.5,2.3);
  \draw[dashed] (0,4.2) -- (1,4.2);

  \draw[>=latex,thick,<->] (0.5,0) -- (2.0,2.3);
  \draw[>=latex,thick,<->] (2.0,2.3) -- (0.5,4.2);
  \draw[>=stealth,<->] (1.6,2) -- (1.6,2.3) node at (1.4, 2.15) {$\Delta$};
  \draw[>=stealth,->] (0.9,4.4) -- (0.9,4.2);
  \draw[>=stealth,->] (0.9,3.8) -- (0.9,4);
  \node at (1.1,4.3) {$\delta$};

\end{tikzpicture}
    \caption{$5\: ^{1}S_{0} - 6\: ^{1}S_{0}$ Raman-type transition.}
    \label{Rabi}
\end{figure}
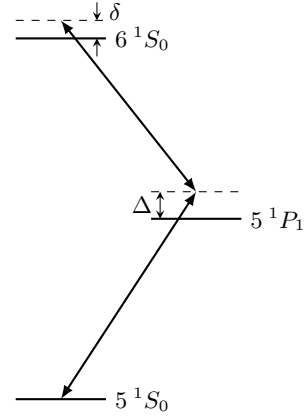

We adiabatically eliminate the intermediate states, and reduce the three level system to a two level system (see Fig.~\ref{Rabi}). 
The two-photon Rabi frequency is
\begin{equation}
    \Omega = \frac{\Omega_1 \Omega_2}{2 \Delta},
\end{equation}
where $\Omega_{1,2}$ are the single-photon Rabi frequencies, and $\Delta$ is the detuning from the $5\: ^1 P_1$ state. We neglect the coupling through other intermediate $P$-states.

The population of the two-photon resonance state $6 \:^1 S_0$ is estimated as~\cite{foot2005atomic}
\begin{equation}
    \rho_{ee} = \frac{\Omega^2/2}{\Omega^2 + \frac{2\Gamma_r}{\Gamma^{'}_r}(\delta^2+\Gamma^{'2}_r/4)},
    \label{excited_state_population}
\end{equation}
where $\delta$ is the two-photon detuning, $\Gamma_r$ is the total population decay rate and $\Gamma^{'}_r$ includes the pressure broadening effect.

\begin{figure}
    \centering
    \includegraphics{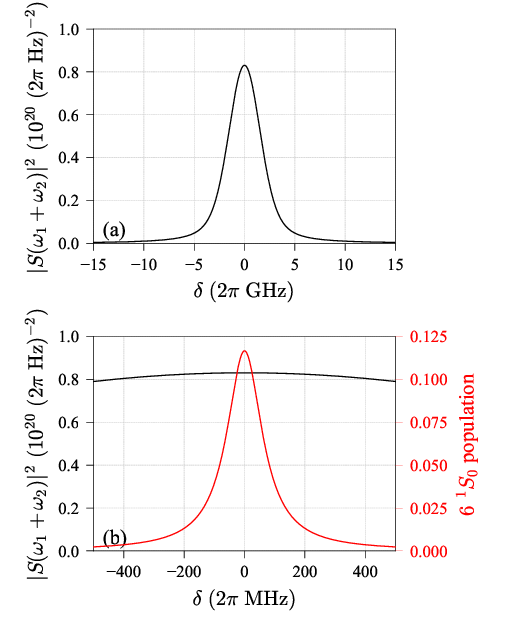}
    \caption{(a) The calculated $\left| S(\omega_1+\omega_2) \right| ^2$ with $5\: ^{1}S_{0}$ as the two-photon resonance state. The linewidth is $2\pi \times 4 \,\mathrm{GHz}$, dominated by the Doppler broadening. (b) The $6\ ^1S_0$ state population as a function of two-photon detuning $\delta$, shown by the red line. It is compared to the ($|S(\omega_1+\omega_2)|^2$), shown by the black line. A detuning of $\delta = 2\pi \times 400\ \mathrm{MHz}$ suppresses the $6\ ^1S_0$ state population without hindering $|S(\omega_1+\omega_2)|^2$.}
    \label{population_delta}
\end{figure}

With 3 W laser power at 375 nm, 6 W at 710 nm, and $b=1.5\ \mathrm{mm}$, we estimate the $6 \:^1 S_0$ excited state population to be $12\%$, indicating a severe saturation effect. 
This effect can be eliminated by slightly detuning the lasers from the two-photon resonance, for example, by $\delta = 2 \pi \times 400\ \mathrm{MHz}$. With this detuning, the population in the $6 \:^1 S_0$ excited state reduces to a negligible value of $0.36\%$, while $S(\omega_1 + \omega_2)$ remains largely unchanged (see Fig.~\ref{population_delta}).

\subsubsection{Amplified spontaneous emission (ASE) effects}

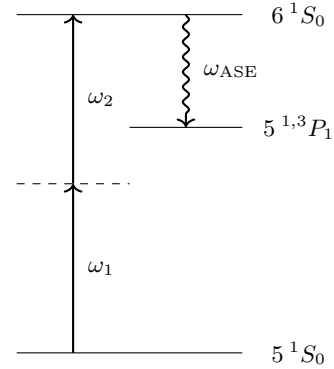
\begin{figure}
    \centering
    \begin{tikzpicture}[scale=1.5]

  \draw (0,0) -- (2,0) node at (2.5,0) {$5\:^{1}S_0$};
  \draw[dashed] (0,1.5) -- (1,1.5);
  \draw (1,2) -- (2,2) node at (2.5,2) {$5\:^{1,3}P_1$};
  \draw (0,3) -- (2,3) node at (2.5,3) {$6\:^{1}S_0$};

  \draw[->, thick] (0.5,0) -- (0.5,1.5);
  \draw[->, thick] (0.5,1.5) -- (0.5,3);
  \draw[->, thick, decorate, decoration={snake,amplitude=.4mm,segment length=2mm,post length=1mm}] (1.5,3) -- (1.5,2);

  \node at (0.75,0.75) {$\omega_1$};
  \node at (0.75,2.25) {$\omega_2$};
  \node at (1.9,2.5) {$\omega_{\mathrm{ASE}}$};

\end{tikzpicture}
   \caption{Amplified spontaneous emission process.}
\label{ASE}
\end{figure}

The population inversion between the $6 \:^1 S_0$ state and the lower-lying $5\: ^{1,3}P_1$ states can cause amplified spontaneous emission. This is another possible source of saturation, which can be represented by a diagram in Fig.~\ref{ASE}.

For relatively small laser intensities, the ASE broadening can be accounted for by replacing the Doppler width in Eq.~\ref{shape_equation} by~\cite{AVSmith1987}:
\begin{equation}
    \Delta\omega_{\mathrm{D}} \to (\Delta \omega_{\mathrm{D}}^2+\Delta \omega_{\mathrm{ASE}}^{2})^{1/2},
\end{equation}
where $\Delta \omega_{\mathrm{ASE}}$ is the power-broadened width associated with the ASE.

To estimate the possible broadening caused by the ASE process for the $6\: ^{1}S_0$ state, we follow a derivation in Ref.~\cite{svelto2010principles}. For a Lorentzian line, the ASE intensity is given by:
\begin{equation}
    I_{\mathrm{ASE}} = \phi I_{s} \Big( \frac{\Omega}{4 \pi} \Big)  \frac{(G-1)^{3/2}}{(G\ln{G})^{1/2}}.
\end{equation}
Here $I_s$ is the saturation intensity, $\phi$ the fluorescence quantum yield, $\Omega$ the emission solid angle, and $G$ the gain of the active medium. 
We assume the active medium length to be twice the Rayleigh range, and use the parameters in last section to calculate the two-photon Rabi frequency. The power-broadened width associated with the ASE is estimated~\cite{Smith:87} to be $\Delta \omega_{\mathrm{ASE}} \approx 25$ MHz, which is negligible compared to the Doppler width.

\subsubsection{Linear absorption of the beams}\label{Twophoton}

\begin{figure}
    \centering
    \includegraphics{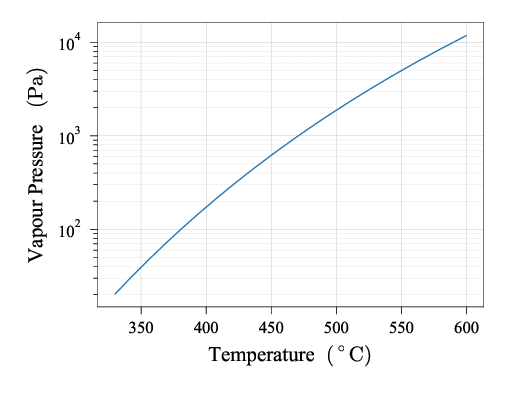}
    \caption{Cadmium vapour pressure. Typically, we get $8500$~$\mathrm{Pa}$ of cadmium at $580$ $\mathrm{^\circ C}$.}
    \label{Cd_VP}
\end{figure}

\begin{figure}
    \centering
    \includegraphics{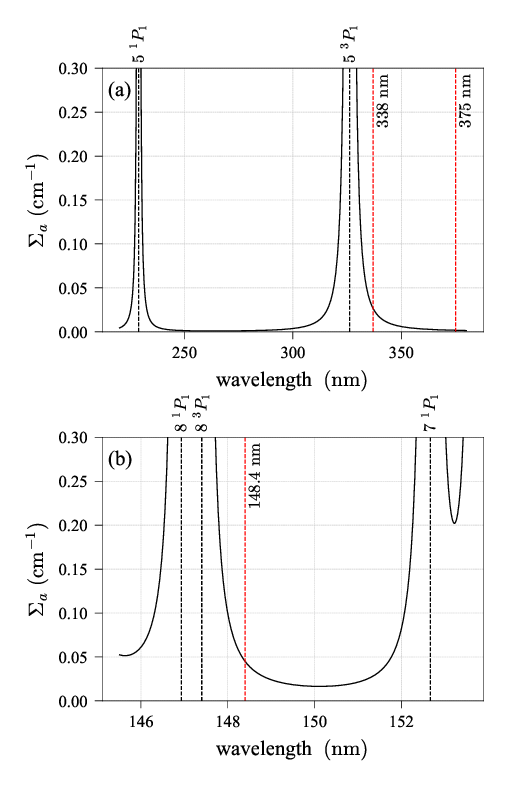}
    \caption{Absorption cross section (in cm$^{-1}$) in the wavelength vicinity of the (a) input beams, (b) output VUV beam, calculated for $N = 7.2 \times 10^{17}$~$\mathrm{cm}^{-3}$, or $T = 580$~$\mathrm{^{\circ} C}$.}
    \label{total_absorption}
\end{figure}

Linear absorption is typically negligible. However, if any of the waves are close to the resonant states, or if the nonlinear medium vapor zone is very long, linear absorption should be considered.
The saturated vapour pressure of cadmium as a function of temperature is shown in Fig.~\ref{Cd_VP}.
The calculated absorption cross section for different wavelengths is plotted in Fig.~\ref{total_absorption}.
The absorption cross section is negligible at 375 nm. 
At 148.4 nm, it amounts to 0.045 cm$^{-1}$, and can be neglected for short medium length.

\subsection{Noble gas for phase matching}\label{RGCompensationSubsec}

At around 148.4~nm, cadmium is negatively dispersive, where the refractive index is smaller at smaller wavelengths. If the required confocal parameter $b$ for a high output VUV power is too small to be practical, 
one could exploit the positive dispersion of noble gases for wave vector mismatch compensation~\cite{AVSmith1987}. 
This extra compensation can also be employed to loose the focus (i.e., enlarge $b$) such that the saturation effect due to the high light intensity in the focus can be mitigated. 
Our calculations show that krypton and xenon gases are good choices for this purpose. 
For the confocal parameter $b=1.5$ mm adopted in the main text, adding a certain amount of krypton gas could improve the VUV power from 30 $\mu$W to 50 $\mu$W.

\subsection{When $\omega_1 \neq \omega_2$}
By tuning $\omega_1$ close to the $5\: ^{1,3}P_1$ resonance states while keeping the two-photon resonance condition ($\omega_1 + \omega_2$ remains unchanged), the output VUV power can be improved.
In this case, the phase matching becomes challenging as the refractive index deviates significantly from unity. The resulting wave vector mismatch can be compensated by lowering the medium's density $N$ or by adding some noble gas. The output VUV power can be enhanced by several times and is highly sensitive to input parameters and medium conditions.

\subsection{Theory of the optical nuclear excitation}

The interaction of the Th-229 nucleus with the VUV light has been treated in~\cite{LarsCalculation}. For simplicity, we consider a nuclear two-level system consisting of a ground state $g$ and an excited state $e$. The laser-induced Rabi frequency $\Omega_{eg}$ can be written as 
\begin{equation}
    \Omega_{eg}=\sqrt{\frac{2 \pi c^2 I_{l} \Gamma_\gamma}{\hbar \omega_0^3}},
    \label{Rabi_frequency}
\end{equation}
where $I_l$ is the laser intensity, $\omega_0$ the angular frequency of the nuclear transition, and $\Gamma_{\gamma}$ the radiative decay rate of the nuclear excited state.

For an extremely low excited state decay rate, as in the case of the Th-229 isomer transition, the condition for the VUV laser linewidth $\Gamma_l$ to observe the Rabi oscillation can be written as 
%Using the definition of Rabi frequency shown in Eq.~\ref{Rabi_frequency}, the Rabi oscillations can be observed if

\begin{equation}
\Gamma_{l} < 2\Omega_{eg} = 2\sqrt{\frac{2 \pi c^2 I_{l} \Gamma_\gamma}{\hbar \omega_0^3}}.
    \label{eq:oscillation conditions}
\end{equation}

\begin{figure}
    \centering
    \includegraphics{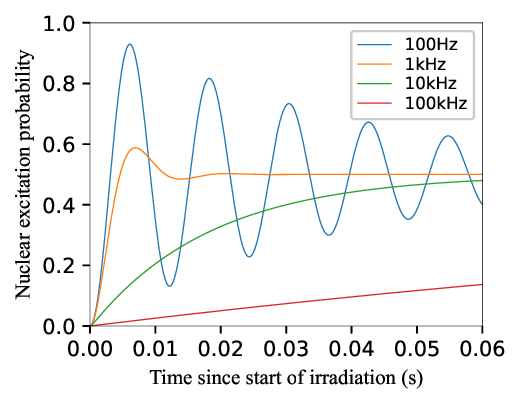}
    \caption{The time-dependent nuclear excitation probability for a single Th-229 nucleus as a function of resonant irradiation time for different laser linewidths.}
    \label{fig:Rabi}
\end{figure}
We assume the VUV laser with a power of $10~\operatorname{\mu W}$ is focused to a spot diameter of 8~$\mathrm{\mu m}$.
To observe the nuclear Rabi oscillation, the VUV laser linewidth needs to satisfy
\begin{equation}
    \Gamma_{l} < 1 \,\mathrm{kHz}.
\end{equation}

The nuclear excitation probability as a function of irradiation time for different laser linewidths is plotted in Fig.~\ref{fig:Rabi}.
As expected, we observe the Rabi oscillation when $\Gamma_l$ is close to or below 1 kHz.


\begin{thebibliography}{67}
\bibitem [{Sup()}]{SupplMat}%
  \BibitemOpen
  \bibinfo {note} {See {S}upplemental {M}aterial at [{URL} will be inserted by publisher], which also includes Refs. [50-56]}\BibitemShut {NoStop}%
\end{thebibliography}
\end{document}